\documentclass[12pt]{article}
\usepackage{amssymb,amsmath,latexsym}
\title{Perturbative test of single parameter scaling \\
for $1D$ random media} 

\author{R. Schrader$^{1}$, H. Schulz-Baldes$^{2}$, A. Sedrakyan$^{3}$
\\
\\
$^1$
{\small  Institut f\"{u}r Theoretische Physik,
Freie Universit\"{a}t Berlin, 14195 Berlin, Germany}
\\$^2$
{\small Institut f\"{u}r Mathematik, Technische Universit\"at Berlin, 
10623 Berlin, Germany}
\\
$^3$
{\small Yerevan Physics Institute,
Yerevan 36, Armenia}
}

\date{ }

\newtheorem{theo}{Theorem}
\newtheorem{defini}{Definition}
\newtheorem{proposi}{Proposition}
\newtheorem{lemma}{Lemma}
\newtheorem{coro}{Corollary}

\newcommand{\CC}{{\mathbb C}}
\newcommand{\NN}{{\mathbb N}}
\newcommand{\RR}{{\mathbb R}}

\newcommand{\ZZ}{{\mathbb Z}}

\newcommand{\Aa}{{\cal A}}

\newcommand{\Pp}{{\cal P}}

\newcommand{\EE}{{\bf E}}
\newcommand{\Bb}{{\cal B}}

\newcommand{\Ss}{{\cal S}}
\newcommand{\Oo}{{\cal O}}
\newcommand{\Tr}{\mbox{\rm Tr}}
\newcommand{\TV}{{\cal T}}

\newcommand{\Qq}{{\cal Q}}
\newcommand{\Kk}{{\cal K}}
\newcommand{\Hh}{{\cal H}}
\newcommand{\Cor}{{\mbox{\rm Cor}}}
\newcommand{\Ad}{{\mbox{\rm Ad}}}
\newcommand{\ad}{{\mbox{\rm ad}}}

\begin{document}

\maketitle

\begin{abstract}
Products of random matrices associated to one-dimensional random media
satisfy a central limit theorem assuring convergence to a gaussian
centered at the Lyapunov exponent.  The hypothesis of 
single parameter scaling states
that its variance is equal to the Lyapunov exponent.  We settle
discussions about its validity for a wide class of models  by proving
that, away from anomalies, 
single parameter scaling holds to lowest order perturbation
theory in the disorder strength. However, it is generically violated
at higher order.  This is explicitely exhibited for the Anderson model.
\end{abstract}

\vspace{.5cm}

\section{Introduction and main result}

One-dimensional quantum systems with a single channel can 
very efficiently be described by $2\times 2$ transfer matrices. 
In a disordered medium,  the transfer matrices are chosen to
be random.
The one-dimensional Anderson model is the proto-type for this class of
models. The most important physical phenomena in these random media
is localization due to multiple coherent wave scattering. It goes
along with positivity of the Lyapunov exponents associated with products
of the random matrices and the Lyapunov exponent is then interpreted
as the inverse localization length of the system. Moreover, the
fluctuations around this asymptotic behavior are gaussian. 
More precisely,
Anderson, Thouless, Abraham and Fisher \cite{ATAF} stated that the 
Landauer conductance follows asymptotically (in the system size) a
log-normal distribution centered at the Lyapunov exponent.
As pointed out by Johnston and Kunz \cite{JK}, this
was a rediscovery of a mathematical result by Tutubalin \cite{Tut} 
(refined by Le Page \cite{LeP}).
The paradigm of single parameter scaling \cite{AALR,ATAF} is then that
there is only one parameter describing the asymptotic behavior of the
random system. For a one-dimensional model this means that the
Lyapunov exponent and the variance of the gaussian should be in some
relation and in fact simply be equal \cite{ATAF}.
The validity of single parameter scaling in this sense
has been analyzed in 
various particular situations \cite{ATAF,SAJ,CRS,DLA,ST}.
The main result of the present 
work can roughly be resumed as follows:
in a wide class of one-dimensional random models
single parameter scaling is valid only perturbatively in a weak disorder
regime and never holds
in a strict sense (exceptional parameter values excluded).

\vspace{.2cm}

For this purpose, we study a general class of one-parameter families
of random transfer matrices exhibiting a so-called critical
energy. This parameter is the effective size of the randomness. In the
Anderson model, it is the coupling constant of the disordered
potential while in the random dimer model \cite{DWP} it is the
distance in energy from the critical energy. We then develop a rigorous
perturbation theory in this parameter. In case of the Lyapunov
exponent, we do not appeal to the random phase approximation
\cite{ATAF,SAJ}, but rather show that phase correlations give a
contribution to lowest order perturbation theory (which, however,
vanishes for the Anderson model). This generalizes
arguments of \cite{Tho,PF,JSS}. On the other hand, the perturbation
theory for the variance is new to our best knowledge. One has to
sum up the phase correlation decay by using adequate counter
terms. This rigorous analysis is made possible by a result of Le Page
(see Section \ref{sec-proofs}). Furthermore, we extend the techniques
of \cite{SS} in order to calculate the scaling exponent of the expectation
value of the Landauer conductance (this is sometimes also called a
generalized Lyapunov exponent). This generalizes results of \cite{Mol}. 

\vspace{.2cm}

Comparing the coefficients to lowest (namely second) order
perturbation theory away from Kappus-Wegner type
anomalies \cite{KW},  we obtain that
the Lyapunov exponent and the variance are equal while the scaling
exponent of the averaged Landauer conductance is twice this value, just
as predicted by \cite{ATAF}. Calculation of higher orders is
possible, but cumbersome in general. For the Anderson model it
becomes feasible and is carried out in Section \ref{sec-Anderson}. We
obtain that the next (namely forth) order contributions are {\it not}
the same for the Lyapunov exponent and the variance and as a consequence 
they are not equal (but close) in the regime of weak disorder. 
For the regime of strong disorder, even 
large discrepancies have been observed
numerically \cite{SAJ}. Deviations from 
single parameter scaling even to lowest order were exhibited at the
band center of the Anderson model, which is the prime example of a
Kappus-Wegner anomaly \cite{ST}. The Lloyd model analyzed in
\cite{DLA} does not fit in our framework because there
the random variables do not have finite moments.
The example of the Anderson model
also allows to show that there does not exist a universal analytic 
function expressing the variance in terms of the Lyapunov exponent.

\vspace{.2cm}

After this brief introduction, let us describe our results more
precisely. The transfer matrices are supposed to be elements of the
following subgroup of the general linear group ${\rm Gl}(2,\CC)$:

\begin{equation}
\label{eq-u11symp}
{\rm U}(1,1)
\;=\;
\{T\in\mbox{Mat}_{2\times 2}(\CC)\;|\;T^*J
T=J\}
\mbox{ , }
\qquad
J
\;=\;
\left(\begin{array}{cc} 0 & -1 \\ 1 & 0 \end{array}\right)
\mbox{ . }
\end{equation}

\noindent Using the conjugation

$$
C^*JC\;=\;\imath \;\Gamma
\mbox{ , }
\qquad
C\;=\;
\frac{1}{\sqrt{2}}\;
\left(\begin{array}{cc} \imath & \imath \\ 1 & -1 \end{array}\right)
\mbox{ , }
\qquad
\Gamma
\;=\;
\left(\begin{array}{cc} 1 & 0 \\ 0 & -1 \end{array}\right)
\mbox{ . }
$$

\noindent one sees that ${\rm U}(1,1)$ is also isomorphic to the
subgroup of matrices $T\in{\rm Gl}(2,\CC)$ satisfying $T^*\Gamma T=\Gamma$.
This representation appears in some applications, but for our purposes
it is more convenient to work with (\ref{eq-u11symp}) because it
contains the standard rotation matrix and the real subgroup
${\rm SL}(2,\RR)$.

\vspace{.2cm}

We will study families
$(T_{\lambda,\sigma})_{\lambda\in\RR,\sigma\in\Sigma}$ of transfer
matrices in ${\rm U}(1,1)$ depending on 
a random variable $\sigma$ in some probability space
$(\Sigma,{\bf p})$ as well as a real coupling parameter
$\lambda$. The dependence on $\lambda$ is supposed to be
smooth.

\begin{defini}
\label{def-critical}
The value $\lambda=0$ is a critical
point of the family 
$(T_{\lambda,\sigma})_{\lambda\in\RR,\sigma\in\Sigma}$ if for all
$\sigma,\sigma'\in\Sigma$:

\begin{equation}
\label{eq-hyp}
\mbox{\rm (i) }\;
[\,T_{0,\sigma},T_{0,\sigma'}]\;=\;0
\mbox{ , }
\qquad
\mbox{\rm (ii) }\;
|\Tr(T_{0,\sigma})|\;<\;2
\mbox{ . }
\end{equation}

\end{defini}

Critical points appear in many applications like in the Anderson model and the
random dimer model \cite{DWP,JSS,Sed}, but also continuous random
Schr\"odinger operators where the transfer matrix is calculated from
a single-site $S$-matrix \cite{KS}. Condition (i) assures that there is
no non-commutativity at $\lambda=0$ (even though the matrices may be
random), while by condition (ii)
the matrices $T_{0,\sigma}$ are conjugated to rotations so that 
there is no {\sl a priori} hyperbolicity in the system.
The example of the Anderson model is studied in more detail in Section
\ref{sec-Anderson}. 

\vspace{.2cm} 

Associated to a given semi-infinite code
$(\sigma_n)_{n\geq 1}$  is
a sequence of matrices $(T_{\lambda,\sigma_n})_{n\geq 1}$. Codes are
random and chosen independently according to the product law ${\bf
p}^{\otimes \NN}$. Averaging w.r.t. ${\bf p}^{\otimes \NN}$ will 
be denoted by $\EE$.
We will suppose that all up to the 5th moment of ${\bf p}$ exist.
In order to shorten notations, we will also write
$T_{\lambda,n}$ for $T_{\lambda,\sigma_n}$. Of interest is the
asymptotic behavior of the random products

$$
\TV_\lambda(N)
\;=\;
\prod_{n=1}^N \,T_{\lambda,n}
\mbox{ . }
$$

\noindent It is first of all characterized by the Lyapunov exponent

\begin{equation}
\label{eq-lyap}
\gamma(\lambda)
\;=\;
\lim_{N\to\infty}
\frac{1}{N}
\,\EE\,\log\left(\left\|\TV_\lambda(N)
\right\|\right)
\mbox{ . }
\end{equation}

\noindent The central limit theorem for products of random matrices 
now states that 

\begin{equation}
\label{eq-LePage}
\frac{1}{\sqrt{N}}
\,\left(\,\log(\|\TV_\lambda(N)e\|)-N\gamma(\lambda)\,\right)
\;\stackrel{{\textstyle \longrightarrow}}{{\scriptscriptstyle
N\to\infty}} \;  
G_{\sigma(\lambda)} 
\end{equation}

\noindent where $G_{\sigma(\lambda)}$ is the centered Gaussian law of
variance $\sigma(\lambda)$ and the convergence is in distribution
independently of the initial unit vector $e$. This was first proven by
Tutubalin under the hypothesis that the measure ${\bf p}$ has a
density \cite{Tut}. Le Page then proved it for arbitrary measures
${\bf p}$ \cite{LeP}. Both proofs can be found in \cite{BL}.
As already discussed above, the single parameter scaling
assumption  is the equality
$\sigma(\lambda)=\gamma(\lambda)$ \cite{ATAF,CRS}. 
Apart from the Lyapunov exponent 
$\gamma(\lambda)$ and the variance $\sigma(\lambda)$,
we are going to analyse the growth exponent of 
the average of the Landauer conductance defined by

\begin{equation}
\label{eq-land}
\hat{\gamma}(\lambda)
\;=\;
\lim_{N\to\infty}
\frac{1}{2N}
\,\log\left(\EE\,\Tr(\TV_\lambda(N)^*\TV_\lambda(N))\right)
\mbox{ . }
\end{equation}

\noindent It follows immediately from Jensen's inequality that 
$\hat{\gamma}(\lambda)\geq \gamma(\lambda)$. Our main results are
resumed in the following:

\begin{theo}
\label{theo-main}
Introduce  the phase $\eta_\sigma\in[0,2\pi)$ 
by $\cos(\eta_\sigma)=\frac{1}{2}\,
\Tr(T_{0,\sigma})$. Suppose 
$\EE(e^{2\imath j\eta_\sigma})\neq 1$ for $j=1,2,3$.
Then there is a constant $D\geq 0$, given in equation {\rm
(\ref{eq-lyapasymp})} below, such that near
the critical point,

\begin{equation}
\label{eq-asymp}
\gamma(\lambda)
\;=\;
D\,\lambda^2\;+\;\Oo(\lambda^3)
\mbox{ , }
\qquad
\hat{\gamma}(\lambda)
\;=\;
2\,D\,\lambda^2\;+\;
\Oo(\lambda^3)
\mbox{ . }
\end{equation}

\noindent If moreover $D>0$, then

\begin{equation}
\label{eq-asymp2}
\sigma(\lambda)\;=\;
D\,\lambda^2\;+\;\Oo(\lambda^3)
\mbox{ . }
\end{equation}

\end{theo}

In Section \ref{sec-lyap}, we also give criteria insuring that $D>0$.
The remainder of the paper contains the proof of this theorem as well
as an analysis of higher orders for the Anderson model.

\vspace{.2cm}

\noindent {\bf Acknowledgments:} This work profited from financial
support of the SFB 288. H. S.-B. would like to thank T. Kottos for
pointing out references \cite{DLA} and \cite{ST}.

\section{Normal form of transfer matrices near critical point}
\label{sec-norm}

If $T\in {\rm U}(1,1)$, then 
$T^*JT=J$ implies that $\det(T)=e^{2\imath\xi}$ for some
$\xi\in[0,\pi)$. Hence $e^{-\imath\xi}T\in {\rm SU}(1,1)
=\{T\in{\rm U}(1,1)|\det(T)=1\}$. 
This means ${\rm U}(1,1)={\rm U}(1)\times {\rm SU}(1,1)$.
Note also that $\Tr(T)=\det(T)\overline{\Tr(T)}$ for 
$T\in {\rm U}(1,1)$ so that
$\Tr(T)\in\RR$ for $T\in{\rm SU}(1,1)$.
One easily verifies that this implies
${\rm SU}(1,1)={\rm SL}(2,\RR)$ is 
a real subgroup of ${\rm Gl}(2,\CC)$. Its Lie algebra is well known:

$$
{\rm sl}(2,\RR)
\;=\;
\left\{
\left.
\left(\begin{array}{cc} a & b \\ c & -a \end{array}\right)
\;\right|
\;
a,b,c\in\RR
\right\}
\mbox{ . }
$$

\vspace{.2cm}

The
eigenvalues of $T\in\,{\rm SL}(2,\RR)$ always come in pairs
$\kappa,1/\kappa$ where 
$\kappa=\mbox{Tr}(T)/2+\imath
\sqrt{1-\mbox{Tr}(T)^2/4}$
and are hence complex conjugate of each other if 
$|\,\Tr(T)|<2$ (which is the case of all transfer matrices near a
critical energy). In this situation, 
the matrix $\mbox{diag}(\kappa,1/\kappa)$ is not in
${\rm SL}(2,\RR)$, however, the associated rotation matrix 
$R_\eta=
\left(\begin{array}{cc} \cos(\eta) & -\sin(\eta) \\ 
\sin(\eta) & \cos(\eta) \end{array}\right)$ with 
$\kappa=e^{\imath \eta}$ is in ${\rm SL}(2,\RR)$
and can be attained with an adequate
conjugation. Hence for any $T\in\,{\rm
U}(1,1)$ with $|\Tr(T)|<2$, 
there exists $M\in \mbox{SL}(2,\RR)$ and $\eta$ 
such that $MTM^{-1}= 
e^{\imath\xi}R_\eta$.

\vspace{.2cm}

Let us now consider the family
$(T_{\lambda,\sigma})_{\lambda\in\CC,\sigma\in\Sigma}$ in ${\rm U}(1,1)$
satisfying
(\ref{eq-hyp}). Because they commute and have a trace less than 2 for
$\lambda=0$, 
they can simultaneously be conjugated to a rotation at that
point. Using a Taylor expansion in $\lambda$, we therefore obtain the
following:

\begin{equation}
MT_{\lambda,\sigma}M^{-1}
\;=\;
e^{\imath \xi_{\sigma}(\lambda)}
\;R_{\eta_\sigma}\;\exp\left(\lambda P_\sigma+\lambda^2 Q_\sigma
+\Oo(\lambda^3)\right)
\mbox{ . }
\label{eq-normal}
\end{equation}

\noindent Here $e^{2\imath \xi_{\sigma}(\lambda)}=\det
(T_{\lambda,\sigma})$ so that $\lambda P_\sigma+\lambda^2
Q_\sigma\in\mbox{sl}(2,\RR)$ for all $\lambda\in\RR$. In particular,
$\Tr(P_\sigma)=\Tr(Q_\sigma)=0$. The constant $C$ in
(\ref{eq-asymp}) only depends on the rotation angles 
$\eta_\sigma$ and on $P_\sigma$ through the constant

\begin{equation}
\label{eq-coeff}
\beta_\sigma
\;=\;
\langle \overline{v}|P_\sigma|v\rangle
\mbox{ , }
\qquad
v\;=\;
\frac{1}{\sqrt{2}}
\left(
\begin{array}{c}
1 \\
-\imath
\end{array}
\right)
\mbox{ . }
\end{equation}

\noindent Note that 
$\beta_\sigma=\langle \overline{v}|P_\sigma^*|v\rangle$,
$|\beta_\sigma|^2=\frac{1}{4}\,\Tr(|P_\sigma|^2+P_\sigma^2)$ and that
$P^2_\sigma=(P_\sigma^*)^2$ is a multiple of the identity. Moreover,
$v$ and $\overline{v}$ are the eigenvectors of all
rotations $R_\sigma$. 

\vspace{.2cm}

Given $\sigma_n$, let us now
denote the associated random phases, rotations and perturbations by
$\xi_{n}(\lambda),\eta_n,R_n,P_n,Q_n$, thus suppressing the
dependence on the random variable $\sigma_n$. 

\vspace{.2cm}

In order to use the normal form (\ref{eq-normal}) for the calculation
of the Lyapunov and Landauer exponents, let us insert $M^{-1}M$ in
between each pair of transfer matrices. As it only gives boundary
contributions, one may also insert an $M$ to the left and an $M^{-1}$ to
the right. Hence, 

$$
\gamma(\lambda)
\;=\;
\lim_{N\to\infty}
\frac{1}{N}
\,\EE\,\log\left(\left\|
\prod_{n=1}^N \,MT_{\lambda,n}M^{-1}\right\|\right)
\mbox{ , }
$$

\noindent as well as a similar formula for 
$\hat{\gamma}(\lambda)$.
Now it is clear that there is no use in carrying along 
the phases $e^{\imath
\xi_{\sigma}(\lambda)}$ in (\ref{eq-normal}) if one is interested in
calculating $\gamma(\lambda)$ and $\hat{\gamma}(\lambda)$.
Therefore we may set from now on
$\xi_{\sigma}(\lambda)=0$. This is equivalent to
supposing that $T_{\lambda,\sigma}\in \mbox{SL}(2,\RR)$.
Moreover, one may factor out the subgroup $\{{\bf 1},-{\bf 1}\}$,
namely even work with the projection of
$MT_{\lambda,\sigma}M^{-1}$ into 
$\mbox{PSL}(2,\RR)=\mbox{SL}(2,\RR)/\{{\bf 1},-{\bf 1}\}$.

\section{Further preliminaries and notations}
\label{sec-notations}

Unit vectors in $\RR^2$ will be denoted by:

\begin{equation}
e_\theta
\;=\;
\left(
\begin{array}{cc} \cos(\theta) \\ \sin(\theta)
\end{array}
\right)
\mbox{ , }
\qquad
\theta\in [0,2\pi)
\mbox{ . }
\label{eq-pruefer}
\end{equation}

\noindent Each transfer matrix $T_{\lambda,\sigma}$ induces an action
on unit vectors via

\begin{equation}
e_{\Ss_{\lambda,\sigma}(\theta)}
\;=\;
\frac{MT_{\lambda,\sigma}M^{-1}e_{\theta}}{
\|MT_{\lambda,n}M^{-1}e_{\theta}\|}
\mbox{ . }
\label{eq-action}
\end{equation}

\noindent Using the vector $v$ of (\ref{eq-coeff}), this is equivalent
to 

\begin{equation}
\label{eq-phasegeneral}
e^{2\imath \Ss_{\lambda,\sigma}(\theta)}
\; =\;
2\,\frac{\langle v|MT_{\lambda,\sigma}M^{-1}|e_\theta\rangle^2}{
\|MT_{\lambda,\sigma}M^{-1}e_\theta\|^2}
\; =\;
\frac{\langle v|MT_{\lambda,\sigma}M^{-1}|e_\theta\rangle}{
\langle \overline{v}|MT_{\lambda,\sigma}M^{-1}|e_\theta\rangle}
\mbox{ . }
\end{equation}

\noindent Now given an initial condition
$\theta_0$, a random
sequence of phases $\theta_n$ associated to a code
$(\sigma_n)_{n\in\NN}$ is iteratively defined by

\begin{equation}
\theta_{n}\;=\;
\Ss_{\lambda,\sigma_n}(\theta_{n-1})
\mbox{ . }
\label{eq-action2}
\end{equation}

\noindent A probability measure $\nu$ on $S^1$ is called invariant for
this random dynamical system if

$$
\int d\nu(\theta)\,f(\theta)
\;=\;
\int d\nu(\theta)\,\EE_\sigma\,f(\Ss_{\lambda,\sigma}(\theta))
\mbox{ , }
\qquad
f\in C(S^1)
\mbox{ , }
$$

\noindent where $\EE_\sigma$ denotes the average w.r.t. ${\bf p}$.
Due to a theorem of Furstenberg \cite{BL} $\nu$ exists and is unique
whenever the Lyapunov exponent is positive. Positivity of the Lyapunov
exponent for $\lambda\neq 0$ is guaranteed by condition (iii) of Definition
\ref{def-critical} (we leave it to the reader to verify that the subgroup
generated by the transfer matrices $T_{\lambda,\sigma}$ is then
non-compact so that Furstenberg's criterium is satisfied \cite{BL}).
In the sequel, $\EE_\nu$
will mean averaging w.r.t. $\nu$ as well as 
the whole code $(\sigma_n)_{n\in\NN}$.

\vspace{.2cm}

Next let us turn to the Lyapunov exponent. According to \cite[A.III.3.4]{BL}
it is given by

%

\begin{equation}
\label{eq-lyap2}
\gamma(\lambda)
\;=\;
\lim_{N\to\infty}
\frac{1}{N}\,\EE\,\log\left(\left\|
\prod_{n=1}^N \,MT_{\lambda,n}M^{-1}e_\theta\right\|\right)
\mbox{ . }
\end{equation}

\noindent As one can also insert the invariant measure, it is also given by
the so-called Furstenberg formula:

$$
\gamma(\lambda)
\;=\;
\int d\nu(\theta)\,\EE_\sigma\,\log\left(\left\|
MT_{\lambda,\sigma}M^{-1}e_\theta\right\|\right)
\mbox{ . }
$$

\noindent Finally, one can
use the random phase dynamics (\ref{eq-action2}) in order to rewrite 
(\ref{eq-lyap2})  as

\begin{equation}
\label{eq-lyap3}
\gamma(\lambda)
\;=\;
\lim_{N\to\infty}
\frac{1}{N}
\;\sum_{n=1}^N
\,\EE_\nu\,\log\left(\left\|
MT_{\lambda,n}M^{-1}e_{\theta_{n-1}}\right\|\right)
\mbox{ . }
\end{equation}

\vspace{.2cm}

\section{Asymptotics of the Lyapunov exponent}
\label{sec-lyap}

Let us  introduce the random variable

\begin{equation}
\label{eq-lyapterm}
\gamma_n
\;=\;
\log\left(\left\|
\,MT_{\lambda,n}M^{-1}e_{\theta_{n-1}}\right\|\right)
\mbox{ . }
\end{equation}

\noindent Then the results cited in the previous section imply

\begin{equation}
\gamma(\lambda)
\;=\;
\lim_{N\to\infty}
\frac{1}{N}\;
\EE_\nu\,\sum_{n=1}^N\,\gamma_n
\;=\;
\lim_{n\to\infty}\EE\,(\gamma_n)
\label{eq-gamma1}
\mbox{ . }
\end{equation}

\noindent Our first aim is to derive a perturbative formula for $\gamma_n$. 
Replacing (\ref{eq-normal}), using that
$R_n$ is orthogonal and expanding the logarithm  shows

\begin{equation}
\label{eq-lyaptermexpan}
\gamma_n
\;=\;
\frac{\lambda}{2}
\langle e_{\theta_{n-1}}|
\tilde{P}_n|e_{\theta_{n-1}}\rangle
+
\frac{\lambda^2}{2}
\langle e_{\theta_{n-1}}|
(\tilde{Q}_n +|P_n|^2+P_n^2)|e_{\theta_{n-1}}\rangle
-
\frac{\lambda^2}{4}
\langle e_{\theta_{n-1}}|
\tilde{P}_n|e_{\theta_{n-1}}\rangle^2
\;+\;\Oo(\lambda^3)
\;.
\end{equation}

\noindent where we have used that
$(P_n^2)^*=P_n^2$ is a multiple of the identity
and set $\tilde{P}_n=P_n+P_n^*$ and 
$\tilde{Q}_n=Q_n+Q_n^*$.
Now for any $T\in
\mbox{Mat}_{2\times 2}(\RR)$, one has

$$
\langle e_\theta|T|e_\theta\rangle
\;=\;
\frac{1}{2}
\,\Tr(T)\;+\;
\Re e 
\left(
\langle \overline{v}|T|v\rangle\,e^{2\imath\theta}
\right)
\mbox{ . }
$$ 

\noindent Hence using the definition (\ref{eq-coeff}) and the remark
following it, we deduce

\begin{equation}
\label{eq-coeffexpan}
\gamma_n
\;=\;
\frac{\lambda^2}{2}\,|\beta_n|^2
+\Re e\left(
\lambda\beta_ne^{2\imath \theta_{n-1}}
-
\frac{\lambda^2}{2}\;\beta_n^2\;
e^{4\imath \theta_{n-1}}
\,+\,\frac{\lambda^2}{2}\,
\langle\overline{v}|(|P_n|^2+\tilde{Q}_n)|v\rangle\,e^{2\imath\theta_{n-1}}
\right)
\;+\;\Oo(\lambda^3)
\;.
\end{equation}

\noindent The so-called random phase approximation consists in
supposing that the angles $\theta_{n-1}$ are distributed according the
Lebesgue measure ({\it i.e.} $\nu$ is the Lebesgue measure). Then only
the non-oscillatory term in (\ref{eq-coeffexpan}) would contribute so
that one would get
$\gamma(\lambda)=\frac{1}{2}\lambda\EE_\sigma(|\beta_\sigma|^2) +
\Oo(\lambda^3)$. In general, however, this is erroneous.
Replacing (\ref{eq-coeffexpan}) into (\ref{eq-lyapterm}), one has to
calculate the following oscillatory sums (as in \cite{JSS}).

\begin{lemma}
\label{lem-osci} 
For $j=1,2$, set

$$
I_j(N)
\;=\;
\EE\;\frac{1}{N}\,
\sum_{n=0}^{N-1} \,e^{2j\imath \theta_n}
\mbox{ . }
$$

\noindent Suppose $\EE_\sigma\left(e^{2j\imath\eta_\sigma}\right)
\,\neq\,1$ for $j=1,2$. Then

$$
I_1(N)
\;=\; 
\frac{\lambda\,\EE_\sigma \left(\overline{\beta_\sigma}\,
e^{2\imath\eta_\sigma}\right)}{1-
\EE_\sigma \left(e^{2\imath\eta_\sigma}\right)}
+
\Oo(\lambda^2,N^{-1})
\mbox{ , }
\qquad
I_2(N)
\;=\;
\Oo(\lambda,N^{-1})
\mbox{ . }
$$

\end{lemma}

\noindent {\bf Proof.} It follows from (\ref{eq-phasegeneral})
and (\ref{eq-normal}) that

\begin{equation}
e^{2\imath \theta_n}
\;=\;
2\,e^{2\imath\eta_n}\,
\frac{\langle v|({\bf 1}+\lambda P_n)|e_{\theta_{n-1}}\rangle^2}{
\langle ({\bf 1}+{\lambda} \,{P_n})
e_{\theta_{n-1}}|({\bf 1}+\lambda P_n)e_{\theta_{n-1}}\rangle}
\;+\;
\Oo(\lambda^2)
\mbox{ . }
\label{eq-phasedev}
\end{equation}

\noindent In particular, this implies

$$
e^{2\imath \theta_n}
\;=\;
e^{2\imath(\eta_n+\theta_{n-1})}
\;+\;
\Oo(\lambda)
\mbox{ , }
$$

\noindent so that, replacing this in each term, one obtains 

$$
I_j(N)
\;=\;
\EE_\sigma\left(e^{2j\imath\eta_\sigma}\right)
\;
I_j(N)
\;+\;
\Oo(\lambda,N^{-1})
\mbox{ . }
$$

\noindent The induction hypothesis therefore implies that 
$I_j(N)=\Oo(\lambda,N^{-1})$. In order to calculate the contribution
of $\Oo(\lambda)$ to $I_1(N)$, let us expand
(\ref{eq-phasedev}). Some algebra shows that 

\begin{equation}
\label{eq-phaseexpan}
e^{2\imath \theta_n}
\,=\,
e^{2\imath(\eta_n+\theta_{n-1})}
\,-\,\lambda\,e^{2\imath(\eta_n+\theta_{n-1})}
\left(
e^{2\imath\theta_{n-1}}
\beta_n
-2\langle v|P_n|v\rangle
-
e^{-2\imath\theta_{n-1}}
\overline{\beta_n}\,
\right)
\;+\;
\Oo(\lambda^2)
\mbox{ .}
\end{equation}

\noindent From the last three terms, those containing still an
oscillatory factor $e^{2\imath\theta_n}$ or $e^{4\imath\theta_n}$
will not contribute to leading order $\lambda$ 
due to the above. Thus we deduce

$$
I_1(N)
\;=\;
\EE_\sigma\left(e^{2\imath\eta_\sigma}\right)
\;
I_1(N)
\;+\;
\lambda\,\EE_\sigma\left(e^{2\imath\eta_\sigma}
\,\overline{\beta_\sigma}\right)
\;+\;
\Oo(\lambda^2,N^{-1})
\mbox{ . }
$$

\noindent This implies the result.
\hfill $\Box$

\vspace{.2cm}

As $\lim_{N\to\infty} I_j(N)=\EE_\nu(e^{2\imath j \theta})$, the lemma
shows that $\EE_\nu(e^{2\imath \theta})=\Oo(\lambda)$ and calculates
the lowest order contribution. If $\beta_\sigma$ is
centered (as for the Anderson model), one even has
$\EE_\nu(e^{2\imath \theta})=\Oo(\lambda^2)$. This shows in particular 
how far
the invariant measure $\nu$ is away from the Lebesgue measure and
that phase correlations are indeed present.

\vspace{.2cm}

Now we can proceed with the calculation of $\gamma(\lambda)$. 
Carrying out the algebra shows $\gamma(\lambda)=D\,\lambda^2+
\Oo(\lambda^3)$ where

\begin{equation}
\label{eq-lyapasymp}
D
\;=\;
\frac{1}{2}
\,\EE_\sigma(|\beta_\sigma|^2)
\;+\;
\Re e\left(\frac{ \EE_\sigma(\beta_\sigma)\,
\EE_\sigma(\overline{\beta_\sigma}\,e^{2\imath\eta_\sigma})}{
1-\EE_\sigma(e^{2\imath\eta_\sigma})}
\right)
\mbox{ . }
\end{equation}

\noindent The second summand is due to phase correlations. It is
important, {\it e.g.}, in the random polymer model \cite{JSS}.
In the Anderson model treated in Section \ref{sec-Anderson}, phase
correlations only contribute to the forth order in $\lambda$. 
Because $\gamma(\lambda)\geq 0$, the coefficient $D$ defined by
(\ref{eq-lyapasymp}) has to be non-negative. More precisely, we prove: 

\begin{proposi}
\label{prop-pos} 
Suppose $\EE_\sigma\left(e^{2\imath\eta_\sigma}\right)
\,\neq\,1$. Then $D$ is always non-negative. 
$D$ vanishes if and only if one of the following two mutually excluding
cases occurs:

\vspace{.1cm}

\noindent {\rm (i)} Both $e^{2\imath\eta_\sigma}$ and $\beta_\sigma$ 
are ${\bf p}$-a.s. constant.

\vspace{.1cm}

\noindent {\rm (ii)} $\EE_\sigma\left(e^{2\imath\eta_\sigma}\right)=0$ and 
$\beta_\sigma$ is a constant multiple of $\;1-e^{2\imath\eta_\sigma}$.

\end{proposi}

\noindent {\bf Proof.} 
Assume first that $e^{2\imath\eta_\sigma}$ is a.s. constant. 
Then $\EE_\sigma\left(e^{2\imath\eta_\sigma}\right)=
e^{2\imath\eta_\sigma}\neq 1$ and it follows from
$\Re e(1-e^{\imath\varphi})^{-1}=1/2$ that 
$2\,D=\EE_\sigma(|\beta_\sigma|^2)-|\EE_\sigma(\beta_\sigma)|^2$. 
By the Cauchy-Schwarz inequality, $D\geq 0$ and $D=0$ if and only if
$\beta_\sigma$ is a.s. constant.
Now let us assume that $e^{2\imath\eta_\sigma}$ is not a.s. constant so that 
$|\EE_\sigma\left(e^{2\imath\eta_\sigma}\right)|<1$. The proposition then
follows from the following lemma by setting $\Hh=L^2({\bf p})$,
$\psi_1=1$ and $\psi_2=e^{2\imath\eta_\sigma}$.
\hfill $\Box$

\begin{lemma}
\label{lem-form} 
Let $\psi_1$ and $\psi_2$ be two linearly independent unit vectors 
in a Hilbert space $\Hh$, implying $|\langle\psi_1|\psi_2\rangle| <1$.  
Then the quadratic form 

$$
\Qq(\psi)
\;=\;
\langle\psi|\psi\rangle
+
\frac{1}{1-\langle\psi_1|\psi_2\rangle}
\;\langle\psi|\psi_2\rangle\; \langle\psi_1|\psi\rangle
+
\frac{1}{1-\langle\psi_2|\psi_1\rangle}
\;\langle\psi|\psi_1\rangle\;\langle\psi_2|\psi\rangle
$$

\noindent on $\Hh$ is positive semi-definite. 
It is positive definite if and only if 
$\langle\psi_1|\psi_2\rangle \neq 0$. 
If $\langle\psi_1|\psi_2\rangle=0$, then $\Qq(\psi)=0$ 
if and only if $\psi$ is a multiple of $\psi_1-\psi_2$.
\end{lemma}

\noindent {\bf Proof.}
Let $\Kk$ be the two dimensional subspace of $\Hh$ 
spanned by $\psi_1$ and $\psi_2$ and $\Kk^{\perp}$ its orthogonal
complement. For an arbitrary vector $\psi$, we write  
$\psi=\psi^{\prime}+\psi^{\prime\prime}$ with
$\psi^{\prime}\in\Kk$ and $\psi^{\prime\prime}\in 
\Kk^{\perp}$. Then
$\Qq(\psi)=\Qq(\psi^{\prime})+\| \psi^{\prime\prime}\|^2$.
Hence it suffices to assume $\psi=\psi^{\prime}\in\Kk$, 
{\sl i.e.} we may restrict $\Qq$ to $\Kk$.
If we introduce the orthonormal basis 
$$
e_1\;=\;\psi_1
\;,
\qquad 
e_2\;=\;
\frac{1}{\sqrt{1-|\langle\psi_1|\psi_2\rangle|^2}}
\;\left(\psi_2-
\langle\psi_1|\psi_2\rangle\,\psi_1\right)
$$

\noindent in $\Kk$, 
then with respect to this orthonormal basis the quadratic form $\Qq$ 
on $\Kk$ is represented by the $2\times 2$ matrix
$$
{\bf Q}
\;=\;
\left(\begin{array}{cc}
\frac{1+|\langle\psi_1|\psi_2\rangle|^2}{(1-\langle\psi_1|\psi_2\rangle)
(1-\langle\psi_2|\psi_1\rangle)}
&
\frac{\sqrt{1-|\langle\psi_1|\psi_2\rangle|^2}}{1-\langle\psi_2|\psi_1\rangle}
\\
& \\
\frac{\sqrt{1-|\langle\psi_1|\psi_2\rangle|^2}}{1-\langle\psi_1|\psi_2\rangle} &1
\end{array}\right)
\;.
$$
Now its trace is obviously strictly positive, while its determinant satisfies
$$
\det {\bf Q}
\;=\;
\frac{2\,|\langle\psi_1|\psi_2\rangle|^2}{(1-\langle\psi_1|\psi_2\rangle)
(1-\langle\psi_2|\psi_1\rangle)} 
\;\geq\; 0
\;.
$$

\noindent
Hence both eigenvalues of ${\bf Q}$ are strictly positive whenever $\psi_1$
and $\psi_2$ are not orthogonal, 
while one eigenvalue is strictly positive and the other one zero if they are 
orthogonal. For this zero eigenvalue the corresponding eigenvector of ${\bf
Q}$ is proportional to the vector
$\left(\begin{array}{c}1\\-1\end{array}\right)$.
This concludes the proof of the lemma.
\hfill $\Box$

\vspace{.2cm}

\section{Perturbation theory for the variance}
\label{sec-variance}

Using telescoping as in (\ref{eq-gamma1}), 
the variance is given by
\begin{eqnarray}
\sigma(\lambda)
\!\! & = & \!\!
\lim_{N\to\infty}\frac{1}{N}\;
\sum_{n,k=1}^N\,\EE_\nu\,
\Big[(\gamma_n-\gamma)
\,(\gamma_k-\gamma)
\Big]
\nonumber
\\
\!\! & = &\!\!
\lim_{N\to\infty}\frac{1}{N}\;
\sum_{n=1}^N\;
\EE_\nu\!
\left[
\gamma_n^2-\gamma^2
+2\sum_{m=1}^{N-n}
\Big(\gamma_n\gamma_{n+m}-\gamma^2\Big)
\right],
\nonumber
\end{eqnarray}

\noindent where we wrote $\gamma$ for $\gamma(\lambda)$ for notational
simplicity. Let us show that the sum over $m$ is convergent even if
$N\to\infty$. To this aim, we denote by $\EE_n$ the expectation over all
$\sigma_m$ with $m\geq n$ such that with the previous notation  
$\EE_1=\EE$. From the remark following Proposition \ref{prop-LePage} 
in Section \ref{sec-proofs} it
follows that $\EE_n(\gamma_{n+m})$ converges exponentially
fast to $\gamma$ for $m\to\infty$. Moreover,
the summands of the sum over $n$ 
converge in expectation so that, if $\nu$ is the unique
invariant measure (integration w.r.t. $\theta_0$), the variance is given by

\begin{equation}
\label{eq-variancedef}
\sigma(\lambda)
\;=\;
\EE_\nu
\left[
\gamma_1^2-\gamma^2
+2\sum_{m=2}^{\infty}
\left(\gamma_1\gamma_m-\gamma^2\right)
\right]
\;=\;
\EE_\nu
\left[
\gamma_1^2-\gamma^2
+2\,\gamma_1\,\EE_2\,\sum_{m=2}^{\infty}
\left(\gamma_m-\gamma\right)
\right].
\end{equation}

\noindent Hence we need to evaluate the sums appearing in the
following lemma. Its proof is deferred to Section
\ref{sec-proofs}. 

\begin{lemma}
\label{lem-osci2} 
Suppose $\EE_\sigma\left(e^{2\imath j\eta_\sigma}\right)
\,\neq\,1$ for $j=1,2,3$ and $D>0$. Then

$$
\EE_2\;
\sum_{m=2}^{\infty} \,
\left(\gamma_{m}
\,-\,\gamma\right)
\;=\; 
\lambda\;
\Re e\;\frac{\EE_\sigma(\beta_\sigma)\,}{1-
\EE_\sigma \left(e^{2\imath\eta_\sigma}\right)}
\;e^{2\imath(\theta_0 +\eta_1)}
\;+\;
\Oo(\lambda^2)
\mbox{ . }
$$

\end{lemma}

\vspace{.2cm}

As $\gamma=\Oo(\lambda^2)$, the first summand $\gamma^2$ gives no
contribution. Therefore we obtain

$$
\sigma(\lambda)
\;=\;
\EE_\nu
\left[
\gamma_1^2
\,+\,\Re e\;
\frac{2\,\lambda\,\EE_\sigma(\beta_\sigma)}{1-
\EE_\sigma \left(e^{2\imath\eta_\sigma}\right)}
\;\gamma_1\,
e^{2\imath(\theta_0+\eta_1)}
\right]
\;+\;
\Oo(\lambda^3)
\mbox{ . }
$$

\noindent Let us calculate the first contribution supposing that $\nu$ is
the invariant measure. Then extracting the linear coefficient in
$\lambda$ from (\ref{eq-coeffexpan}) and
using Lemma \ref{lem-osci}

\begin{eqnarray}
\EE_\nu\,(
\gamma_1^2)
& = & 
\lim_{N\to\infty}
\,\frac{1}{N}\;\EE_\nu\,\sum_{n=1}^{N}\,\gamma_n^2
\nonumber
\\
& = & 
\frac{\lambda^2}{4}\;
\lim_{N\to\infty}
\,\frac{1}{N}\;\EE_\nu\,\sum_{n=1}^{N}\,
\left(\beta_n\, e^{2\imath\theta_{n-1}}+\overline{\beta_n}
\, e^{-2\imath\theta_{n-1}}\right)^2
\,+\,\Oo(\lambda^3)
\nonumber
\\
& = & 
\frac{\lambda^2}{2}\;\EE_\sigma\left(|\beta_\sigma|^2\right)
\,+\,\Oo(\lambda^3)
\mbox{ . }
\nonumber
\end{eqnarray}

\noindent Similarly

$$
\EE_\nu
\left(
\gamma_1\,e^{2\imath(\theta_0 +\eta_1)}
\right)
\;=\;
\frac{\lambda}{2}\,\EE_\sigma\left(
\overline{\beta_\sigma}\,e^{2\imath\eta_\sigma}
\right)
\,+\,\Oo(\lambda^2)
\mbox{ . }
$$

\noindent This implies

$$
\sigma(\lambda)\;=\;\gamma(\lambda)\,+\,\Oo(\lambda^3)
\mbox{ . }
$$

\section{Estimates on the correlation decay}
\label{sec-proofs}

The main purpose of this section is to prove
Lemma \ref{lem-osci2}. We choose a slightly more general formulation,
however, allowing to treat also other quantities. 
The first aim is to show how the estimates of
\cite[Theorem A.V.2.5]{BL} (due to Le Page \cite{LeP})
can be made quantitative when combined with Lemma \ref{lem-osci}. Throughout
this section, we suppose that $D>0$ and, for sake of notational simplicity,
$\lambda\geq 0$. Let us introduce a distance on $S^1$ by

$$
\delta(\theta,\psi)
\;=\;
\sqrt{1-\langle e_\theta|e_\psi\rangle^2}
\;=\;
\|e_\theta\wedge e_\psi\|
\;,
$$

\noindent where the second norm is in $\Lambda^2\RR^2$. 
For $\alpha\in(0,1]$, the space of H\"older continuous functions $C_\alpha(S^1)$
is given by the continuous functions $f\in C(S^1)$ with finite H\"older
norm $\|f\|_\alpha=\max\{\|f\|_\infty,m_\alpha(f)\}$, where

$$
m_\alpha(f)
\;=\;
\sup_{\theta,\psi\in S^1}\;
\frac{|f(\theta)-f(\psi)|}{\delta(\theta,\psi)^\alpha}
\mbox{ . }
$$

\noindent Then we 
define $\Ss,\Pp:C_\alpha(S_1)\to C_\alpha(S_1)$ by

$$
(\Ss f)(\theta)\;=\;\EE(f(\Ss_{\lambda,\sigma}(\theta)))
\;,
\qquad
(\Pp f)(\theta)\;=\;\int d\nu(\psi)\,f(\psi)
\;.
$$

\noindent 
Here $\Pp$ should be thought of as the projection on the constant
function with value given by the average w.r.t. the unique invariant
measure. Both $\Ss$ and $\Pp$ depend on $\lambda$.

\begin{proposi}
\label{prop-LePage} 
There exists a constant $c,d>0$ such that for $\alpha=d\,\lambda$,

$$
\| (\Ss^N-\Pp)(f)\|_\alpha
\;\leq\;
\|f\|_\alpha\;
e^{-c\,\lambda^3\,N}
\;.
$$

\end{proposi}

\vspace{.2cm}

\noindent {\bf Proof.} Let us introduce $\zeta:\mbox{SL}(2,\RR)\times S^1
\times S^1\to\RR$ by

$$
\zeta(T,(\theta,\psi))
\;=\;
\log\left(
\frac{\delta(\Ss_T(\theta),\Ss_T(\psi))}{\delta(\theta,\psi)}
\right)
\;,
$$

\noindent where $\Ss_T:S^1\to S^1$ is defined as in (\ref{eq-action}) by
$e_{\Ss_{T}(\theta)}=Te_\theta / \|Te_\theta\|$. Then $\zeta$ is a cocycle,
namely it satisfies

$$
\zeta(T'T,(\theta,\psi))
\;=\;
\zeta(T',(\Ss_T(\theta),\Ss_T(\psi)))
\;+\;
\zeta(T,(\theta,\psi))
\;.
$$

\noindent Moreover, one has

$$
\zeta(T,(\theta,\psi))
\;=\;
\log\left(
\frac{\|\Lambda^2 T\,e_\theta\wedge e_\psi\|}{
\|Te_\theta\|\,\|Te_\psi\|\,\|e_\theta\wedge e_\psi\|}
\right)
\;=\;
-\log(\|Te_\theta\|)-\log(\|Te_\psi\|)
\;,
$$

\noindent because $\|\Lambda^2 T\,e_\theta\wedge e_{\theta+\frac{\pi}{2}}\|
=1$. Therefore 
$|\zeta(MT_{\lambda,n}M^{-1},(\theta,\psi))|\leq c_0\lambda$ and the
cocycle property implies

$$
|\zeta(\TV_{\lambda}(n),(\theta,\psi))|
\;\leq\; 
c_1\,\lambda \,n
\;.
$$

\noindent Furthermore, invoking Lemma \ref{lem-osci} shows that

$$
\EE
\big(
\zeta(\TV_{\lambda}(n),(\theta,\psi))
\big)
\;\leq\;
-2\,D\,\lambda^2\,n+\; c_2\,
\left(\lambda^3\,n+\lambda\right)
\;,
$$
 
\noindent where $D>0$ is the coefficient given in (\ref{eq-lyapasymp}).
Defining the sequence of angles $\psi_n$ as in (\ref{eq-action2}), but with
initial condition $\psi_0=\psi$, one can now infer from 
$e^x\leq 1+x+ x^2 e^{|x|}/2$ that

\begin{eqnarray*}
\EE
\left(
\frac{\delta(\theta_n,\psi_n)^\alpha}{\delta(\theta,\psi)^\alpha}
\right)
& \leq &
1\,+\,\alpha \,\EE
\big(
\zeta(\TV_{\lambda}(n),(\theta,\psi))
\big)
\,+\,
\frac{1}{2}\,\alpha^2 \,
\EE\big(
\zeta(\TV_{\lambda}(n),(\theta,\psi))^2
e^{\alpha\,|\zeta(\TV_{\lambda}(n),(\theta,\psi))|}\big)
\\
& \leq &
1-2\,D\,\alpha\,\lambda^2\,n
\,+\,c_2\,\alpha\,(\lambda^3\,n+\lambda)
\,+\,
\frac{1}{2}\,c_1^2 \,(\alpha\lambda n)^2 \;e^{c_1\,\alpha\lambda n}
\;.
\end{eqnarray*}

\noindent Now one has on the one hand,

$$
m_\alpha((\Ss^n-\Pp)(f))
\;=\;\sup_{\theta,\psi\in S^1}
\;\frac{\big|\,\EE\big(f(\theta_n)-f(\psi_n)\big)\big|}{
\delta(\theta,\psi)^\alpha}
\;\leq\;
m_\alpha(f)\;
\sup_{\theta,\psi\in S^1}
\EE
\left(
\frac{\delta(\theta_n,\psi_n)^\alpha}{\delta(\theta,\psi)^\alpha}
\right)
\;,
$$

\noindent and, furthermore, using the invariance of $\nu$ and then
$\delta(\theta,\psi)\leq 1$,

\begin{eqnarray*}
\|(\Ss^n-\Pp)(f)\|_\infty
& = &
\sup_{\theta\in S^1}
\left|
\int d\nu(\psi)\;
\;\EE\big(f(\theta_n)-f(\psi_n)\big)
\right|
\\
& \leq &
m_\alpha(f)\;\sup_{\theta\in S^1}
\int d\nu(\psi)\;
\EE
\left(
\frac{\delta(\theta_n,\psi_n)^\alpha}{\delta(\theta,\psi)^\alpha}
\right)
\;.
\end{eqnarray*}

\noindent Therefore, we can deduce

$$
\|(\Ss^n-\Pp)(f)\|_\alpha
\;\leq\;
m_\alpha(f)\;\left(
1-2\,\alpha\, D\,\lambda^2\,n
\,+\,c_2\,\alpha\,(\lambda^3\,n+\lambda )
\,+\,
c_1^2 \,(\alpha\lambda n)^2 \;e^{c_1\,\alpha\lambda n}
\right)
\;.
$$

\noindent Now we choose $n=[c_3/\lambda]+1$ (as usual $[x]$ denotes the
integer part of $x\in\RR$) and $\alpha=c_4\lambda$ with
adequate $c_3,c_4>0$ and note that $m_{\alpha'}(f)\leq m_\alpha(f)$ for
$\alpha'\leq\alpha$, so that for some $c_5>0$

$$
\|(\Ss^n-\Pp)(f)\|_\alpha
\;\leq\;
\|f\|_\alpha\,(1-c_5\lambda^2)
\;,
\qquad n=\left[\frac{c_3}{\lambda}\right]+1
\;.
$$

\noindent Finally, we note that $\Ss^N-\Pp=(\Ss^n-\Pp)^{N/n}$ 
because $\Ss\Pp=\Pp\Ss=\Pp$ so that iterating the last inequality
completes the proof.
\hfill $\Box$

\vspace{.2cm}

When applied to the  function 
$f(\theta)=\EE_\sigma(\log(\|MT_{\lambda,\sigma}M^{-1}e_\theta\|))$
(which is H\"older continuous for any $\alpha$)
this estimate directly implies that 
$\EE(\gamma_n)$ converges exponentially fast
to $\gamma$ when $n\to\infty$.
Furthermore, using $m_\alpha(f)\leq \|\partial_\theta f\|_{\infty}$, one now
gets: 

\begin{coro}
\label{coro-sum} 
There is a constant $c>0$ such that

$$
\EE\;\sum_{m=1}^\infty
\;\big(f(\theta_m)-\EE_\nu(f(\theta))\big)
\;\leq\;
c\,
\max
\left\{
\|f\|_\infty,
\|\partial_\theta f\|_\infty
\right\}
\;\lambda^{-3}
\mbox{ . }
$$

\end{coro}

\vspace{.2cm}

We will now consider an algebra $\Aa$ of functions which are analytic on
some neighborhood of $\{0\}\times S^1\subset \CC^2$ and have a power series
of the form

\begin{equation}
\label{eq-algebradef}
F(\lambda,z)
\;=\;
\sum_{k\geq 0}\lambda^k\,\sum_{|l|\leq k} F_{k,l}
\,z^{2l}
\;=\;
\sum_{k\geq 0}\lambda^k\,F_k(z)
\mbox{ , }
\end{equation}

\noindent with complex coefficients $F_{k,l}$. The only elementary fact
needed here is that $F\cdot G\in\Aa$ whenever $F,G\in\Aa$. We will mainly be
interested in the values on $\RR\times S^1$ and there also write
$F(\lambda,\theta)=F(\lambda,2^{2\imath\theta})$. 
Let us give two examples for
elements in $\Aa$. One is the dynamics defined in (\ref{eq-action}),
namely $(\lambda,\theta)\mapsto
e^{2\imath(\Ss_{\lambda,\sigma}(\theta)-\theta)}$ is a function in
$\Aa$ because it is a quotient of analytic functions in $(\lambda,\theta)$ 
of the form (\ref{eq-algebradef}). Taking its $j$th power, one obtains
the representation

\begin{equation}
e^{2\imath j\Ss_{\lambda,\sigma}(\theta)}
\;=\;
e^{2\imath j (\theta+\eta_\sigma)}
\,\left(
1+\sum_{k\geq 1}\lambda^k\sum_{|l|\leq k}c^{\,j}_{k,l}(\sigma)\;
e^{2\imath\,l\,\theta}
\right)
\mbox{ , }
\label{eq-actionexpan}
\end{equation}

\noindent for adequate complex coefficients
$c^{\,j}_{k,l}(\sigma)$. Comparing with 
(\ref{eq-phaseexpan}), we see that 
$c^1_{1,1}(\sigma)=-\beta_\sigma$, $c^1_{1,0}(\sigma)=2\,\langle
v|P_\sigma| v\rangle$ and $c^1_{1,-1}(\sigma)=-\overline{\beta_\sigma}$.

\vspace{.2cm}

Our second example is the function $F(\lambda,\theta)=
\EE_\sigma(\log(\|MT_{\lambda,\sigma}M^{-1}e_\theta\|))$. 
With this function, one has
$\EE_m(\gamma_m)=F(\lambda,\theta_{m-1})$ and
$\gamma(\lambda)=\EE_\nu(F(\lambda,\theta))$. This function 
appears in Lemma \ref{lem-osci2}. 

\vspace{.2cm}

Generalizing Lemma \ref{lem-osci2}, we are led to evaluate
(perturbatively in $\lambda$) the summed
up correlation decay for an arbitrary function $F\in\Aa$:

$$
\Cor(F)(\lambda)
\;=\;
\EE_2\,\sum_{m=1}^\infty 
\big(F(\lambda,\theta_m)-\EE_\nu(F(\lambda,\theta))\big)
\mbox{ . }
$$

\noindent From Corollary \ref{coro-sum} follows the {\it a priori}
estimate $|\Cor(F)(\lambda)|\leq C\,\lambda^{-3}$. For simplicity, let
us now calculate $\Cor(F)(\lambda)$ to order $\lambda$, namely
discard terms of order $\Oo(\lambda^2)$. This is
the situation covered by Lemma \ref{lem-osci2}. Consider
$G(\lambda,\theta)=F(\lambda,\theta)-\sum_{k=0}^4 \lambda^k\,F_k(\theta)
=\Oo(\lambda^{5})$ so that $\|\partial_\theta
G(\lambda,\,.\,)\|_\infty= \Oo(\lambda^{5})$. Combined with
Corollary \ref{coro-sum}, it follows:

\begin{eqnarray}
\Cor(F)(\lambda)
& = &
\EE_2\;\sum_{m=1}^\infty 
\left(G(\lambda,\theta_m)
+\sum_{k=0}^4 \lambda^k\,F_k(\theta_m)
-\sum_{k=0}^4 \lambda^k\,\EE_\nu(F_k(\theta))
-\EE_\nu(G(\lambda,\theta))\right)
\nonumber
\\
& = & 
\sum_{k=0}^4 \lambda^k
\;\EE_2\;\sum_{m=1}^\infty 
\big(
F_k(\theta_m)-\EE_\nu(F_k(\theta))\big)
\;+\;\Oo(\lambda^2)
\nonumber
\\
& = & 
\sum_{k=1}^4 \lambda^k
\,\sum_{|l|\leq k} F_{k,l}\;
\EE_2\;\sum_{m=1}^\infty 
\big(e^{2\imath l \theta_m}-\EE_\nu(e^{2\imath l\theta})
\big)
\;+\;\Oo(\lambda^2)
\mbox{ . }
\nonumber
\end{eqnarray}

\noindent The appearing sum over $m$ is finite (again due to Corollary
\ref{coro-sum}) and, moreover,
we can calculate its value perturbatively.

\begin{lemma}
\label{lem-osci3} 
Let 

$$
J_j
\;=\;
\EE_2\;\sum_{m=1}^\infty 
\big(e^{2\imath j \theta_m}-\EE_\nu(e^{2\imath j\theta})
\big)
\mbox{ , }
\qquad
j\in \ZZ\mbox{ . }
$$

\noindent
Suppose $\EE_\sigma \left(e^{2\imath j\eta_\sigma}\right)
\,\neq\,1$ for $j=1,\ldots,4$. Then $J_4=\Oo(\lambda^{-2})$,
$J_3=\Oo(\lambda^{-1})$, $J_2=\Oo(1)$ and 

\begin{equation}
\label{eq-decaysummed}
J_1\;=\;
\frac{1}{1-
\EE_\sigma \left(e^{2\imath\eta_\sigma}\right)}\;
e^{2\imath(\theta_0+\eta_1)}
\;+\;
\Oo(\lambda)
\mbox{ . }
\end{equation}

\end{lemma}

\vspace{.2cm}

\noindent {\bf Proof.} We use
(\ref{eq-actionexpan}) in order to express
$e^{2\imath j \theta_m}$ in terms of $e^{2\imath j
\theta_{m-1}}$, but truncate the expansion at $\Oo(\lambda^K)$. Using
again Corollary \ref{coro-sum}, we deduce

\begin{eqnarray}
J_j
& = &
\EE_2\;\sum_{m=1}^\infty 
\left[
e^{2\imath j (\theta_{m-1}+\eta_m)}
\left(1+
\sum_{k=1}^K \lambda^k
\sum_{|l|\leq k}
\EE_\sigma(c^{\,j}_{k,l}(\sigma))e^{2\imath l \theta_{m-1}}\right)
\right.
\nonumber
\\
& & 
\;\;\;\;\;\;\;\;\;\;\;\;\;
-
\left.
\EE_\sigma \left(e^{2\imath j\eta_\sigma}\right)\,
\EE_\nu \left(e^{2\imath j \theta}
\left(1+
\sum_{k=1}^K \lambda^k
\sum_{|l|\leq k}
\EE_\sigma(c^{\,j}_{k,l}(\sigma))e^{2\imath l \theta}
\right)\right)\right]
\;+\;\Oo(\lambda^{K-2})
\nonumber
\\
& = & 
\EE_\sigma \left(e^{2\imath j\eta_\sigma}\right)
J_j
\;+\;
e^{2\imath j (\theta_0+\eta_1)}
\;+\;
\EE_\sigma \left(e^{2\imath j\eta_\sigma}\right)\,
\sum_{k=1}^K \lambda^k
\sum_{|l|\leq k}
\EE_\sigma(c^{\,j}_{k,l}(\sigma))\,J_{j+l}
\;+\;\Oo(\lambda^{K-2},\lambda)
\nonumber
\\
& = & 
\frac{1}{1-\EE_\sigma \left(e^{2\imath j\eta_\sigma}\right)}
\left[
e^{2\imath j (\theta_0+\eta_1)}
+
\EE_\sigma \left(e^{2\imath j\eta_\sigma}\right)
\sum_{k=1}^K \lambda^k
\sum_{|l|\leq k}
\EE_\sigma(c^{\,j}_{k,l}(\sigma))\,J_{j+l}
\right]
\;+\;\Oo(\lambda^{K-2},\lambda)
\mbox{ ,}
\nonumber
\end{eqnarray}

\noindent where $\Oo(\lambda^{K-2},\lambda)=\Oo(\lambda^{K-2})
+\Oo(\lambda)$ and in the second equality we used the fact
$\EE_\nu(e^{2\imath j \theta})=\Oo(\lambda)$ (due to Lemma
\ref{lem-osci}). 
As we know that $J_0=0$ and that the {\it a priori} estimate 
$J_j=\Oo(\lambda^{-3})$ holds, this
calculation shows that $J_j=\Oo(\lambda^{-2})$ if 
$\EE_\sigma \left(e^{2\imath j\eta_\sigma}\right)\neq 0$. 
The estimate $J_3=\Oo(\lambda^{-1})$ 
now follows by choosing $K=1$ and replacing
$J_4=\Oo(\lambda^{-2})$ and $J_2=\Oo(\lambda^{-2})$.
The same way one deduces $J_2=\Oo(\lambda^{-1})$ and 
$J_1=\Oo(\lambda^{-1})$.
Choosing $K=2$, a similarly argument shows in the next step
$J_2=\Oo(1)$ and $J_1=\Oo(1)$.
In order to establish (\ref{eq-decaysummed}), let us choose $K=3$. The
first term in the last square bracket gives the desired contribution,
while the sum is $\Oo(\lambda)$ because $J_2=\Oo(1)$,
$J_3=\Oo(\lambda^{-1})$ and $J_3=\Oo(\lambda^{-2})$.
\hfill $\Box$

\vspace{.2cm}

In order to calculate $\Cor(F)(\lambda)$, let us recall that $J_0=0$ and
$J_{-j}=\overline{J_j}$. Thus

$$
\Cor(F)(\lambda)
\;=\;
\lambda\,F_{1,1}\,J_1\;+\;\lambda\,F_{1,-1}\,\overline{J_1}
\;+\;\Oo(\lambda^2)
\mbox{ , }
$$

\noindent with $J_1$ given by Lemma \ref{lem-osci3}. For the function 
$F(\lambda,\theta)=
\EE_\sigma(\log(\|MT_{\lambda,\sigma}M^{-1}e_\theta\|))$, equation
(\ref{eq-coeffexpan}) implies
$F_{1,1}=\frac{1}{2}\EE_\sigma(\beta_\sigma)=\overline{F_{1,-1}}$.
This gives Lemma \ref{lem-osci2}.

\section{Some identities linked to the adjoint representation}
\label{sec-adjoint}

The adjoint
representation of $\mbox{SL}(2,\RR)$ on its Lie algebra is defined by
$\Ad_T(t)=TtT^{-1}$,  $t\in\mbox{sl}(2,\RR)$. It leaves invariant the
quadratic form $q(t,s)=\frac{1}{2}\,\Tr(ts)$ of signature $(2,1)$
(note that $q(t,t)=-\det(t)$). A basis $\Bb=\{b_1,b_2,b_3\}$
of $\mbox{sl}(2,\RR)$ is given by

$$
b_1
\;=\;
\left(\begin{array}{cc} 1 & 0 \\ 0 & -1 \end{array}\right)
\mbox{ , }
\quad
b_2
\;=\;
\left(\begin{array}{cc} 0 & 1 \\ 1 & 0 \end{array}\right)
\mbox{ , }
\quad
b_3
\;=\;
\left(\begin{array}{cc} 0 & -1 \\ 1 & 0 \end{array}\right)
\mbox{ . }
$$

\noindent It is orthonormal w.r.t. the
scalarproduct $\langle s|t\rangle=\frac{1}{2}\Tr(s^*t)$. Moreover,
denoting the coordinate map w.r.t. this basis by $K^\Bb:
\mbox{sl}(2,\RR)\to\RR^3$ and the standard scalar product in $\RR^3$
also by $\langle \vec{x}\,|\,\vec{y}\,\rangle$ for
$\vec{x},\vec{y}\in\RR^3$, we have

$$
q(t,s)
\;=\;
\langle K^\Bb(t)|\,\Gamma_{2,1}\,|K^\Bb(s)\rangle
\mbox{ , }
\qquad
\Gamma_{2,1}
\;=\;
\left(
\begin{array}{ccc} 1 & 0 & 0 \\
0 & 1 & 0 \\
0 & 0 & -1 
\end{array}
\right)
\mbox{ . }
$$

\noindent Finally, let us set $\Ad^\Bb_T=K^\Bb\Ad_T
(K^\Bb)^{-1}$. This means $(\Ad^\Bb_T)_{j,k}=\langle
b_j|\Ad_T(b_k)\rangle$. Note that $J=b_3\in \mbox{SL}(2,\RR)$
and that $\Ad^\Bb_J=-\Gamma_{2,1}$. Thus $T^*JT=J$ implies 
$(\Ad^\Bb_T)^*\Gamma_{2,1}\Ad^\Bb_T=\Gamma_{2,1}$. Hence $\Ad^\Bb_T$ is an
element of the Lorentz group $\mbox{SO}(2,1)$ given by all $A\in
\mbox{Mat}_{3\times 3}(\RR) $ satisfying 
$A^*\Gamma_{2,1} A=\Gamma_{2,1}$. As 
$\Ad^\Bb_{\bf 1}=\Ad^\Bb_{-{\bf 1}}={\bf 1}$, the
adjoint representation gives an isomorphism 
$\mbox{PSL}(2,\RR)\cong\mbox{SO}(2,1)$. Important for the sequel is
that the eigenvalues $\mu_1,\mu_2,\mu_3$ of $A\in
\mbox{SO}(2,1) $ satisfy $\mu_1\mu_2\mu_3=\pm 1$. One of
these eigenvalues, say $\mu_3$, must be real, while the other two may
either be real as well or be complex conjugates of each other.

\vspace{.2cm}

Next it follows from a short calcuation that for 
$T=\left(\begin{array}{cc} \hat a & \hat b \\ \hat c & \hat d \end{array}\right)
\in\mbox{SL}(2,\RR)$

$$
\Ad^\Bb_T
\;=\;
\left(
\begin{array}{ccc} 
\hat a\hat d+\hat b\hat c & \hat d\hat b-\hat a\hat c & \hat a\hat c+\hat b\hat d \\  
\hat d\hat c-\hat a\hat b & \frac{1}{2}(\hat d^2-\hat b^2-\hat c^2+\hat a^2) & \frac{1}{2}(\hat d^2-\hat b^2+\hat c^2-\hat a^2) \\
\hat d\hat c+\hat a\hat b & \frac{1}{2}(\hat d^2+\hat b^2-\hat c^2-\hat a^2) & \frac{1}{2}(\hat d^2+\hat b^2+\hat c^2+\hat a^2) 
\end{array}
\right)
\mbox{ . }
$$

\noindent In particular, for a rotation $R_\eta$ by $\eta$ we get

$$
\Ad^\Bb_{R_\eta}
\;=\;
\left(
\begin{array}{ccc} 
\cos(2\eta) & -\sin(2\eta) & 0 \\  
\sin(2\eta) & \cos(2\eta) & 0 \\
0 & 0 & 1 
\end{array}
\right)
\mbox{ . }
$$

\noindent The eigenvalues are $e^{2\imath\eta},\,
e^{-2\imath\eta},\,1$
with respective eigenvectors $\vec{v}_1=(\vec{e}_1-\imath\vec{e}_2)/\sqrt{2},\,
\vec{v}_2=(\vec{e}_1+\imath\vec{e}_2)/\sqrt{2},
\,\vec{v}_3=\vec{e}_3$. 

\vspace{.2cm}

We will also need the adjoint representation of the normal form
(\ref{eq-normal}):

\begin{equation}
\label{eq-adjointperturb}
\Ad_{R\,\exp(\lambda P +\lambda^2 Q)}
\;=\;
\Ad_{R}
\left({\bf 1}\;+\;
\lambda\,\ad_{P}
\;+\;\lambda^2\,\ad_{Q}\;+\;\frac{\lambda^2}{2}\,(\ad_{P})^2
\;+\;\Oo(\lambda^3)
\right)
\mbox{ , }
\end{equation}

\noindent where $\ad_t(s)=[t,s]$ 
for $t,s\in \mbox{sl}(2,\RR)$.
Let us write out more explicit formulas in the representation 
w.r.t. the basis $\Bb$. 
For $P=\left(\begin{array}{cc} a & b \\ c & -a \end{array}\right)
\in\mbox{sl}(2,\RR)$,

\begin{equation}
\label{eq-adform}
\ad^\Bb_P
\;=\;
\left(
\begin{array}{ccc} 
0 & b-c & b+c \\  
c-b & 0 & -2a \\
b+c & -2a & 0
\end{array}
\right)
\mbox{ , }
\end{equation}

\noindent and

\begin{equation}
\label{eq-bdform}
(\ad^\Bb_P)^2
\;=\;
\left(
\begin{array}{ccc} 
4bc & -2a(b+c) & 2a(c-b) \\  
-2a(b+c) & 4a^2-(c-b)^2 & c^2-b^2 \\
2a(b-c) & b^2-c^2 & (b+c)^2+4a^2
\end{array}
\right)
\mbox{ . }
\end{equation}

\vspace{.2cm}

Finally let ${\cal P}:\CC^2\otimes\CC^2\to\CC^2\otimes\CC^2$ be the permutation
operator ${\cal P}\phi\otimes\psi=\psi\otimes\phi$. It can readily be checked
that 

$$
{\cal P}
\;=\;
\frac{1}{2}
\left(
{\bf 1}\otimes{\bf 1}
\;-\;
\sum_{j=1}^3\,\det(b_j)\;b_j\otimes b_j
\right)
\mbox{ . }
$$

\noindent Multiplying this identity from the left by $T\otimes J$ and
from the right by ${\bf 1}\otimes T^{-1}J$, one gets for
$T\in\mbox{SL}(2,\RR)$,

\begin{equation}
\label{eq-identity}
T\otimes T^t\,{\cal P}
\;=\;
\frac{1}{2}
\left(\,-\,
J\otimes J
\;+\;
\sum_{j,k=1}^3\,\det(b_j)\;(\Ad_T^\Bb)_{k,j}\; b_kJ\otimes Jb_j
\right)
\mbox{ . }
\end{equation}

\noindent This is useful for the calculation of the Landauer
conductance because

$$
\|T w\|^2
\;=\;
\langle w|\,\Tr_1 (T\otimes T^t\,{\cal P})|w\rangle
\mbox{ , }
$$

\noindent where $\Tr_1$ is the partial trace over the first component
of $\CC^2\otimes\CC^2$. Replacing (\ref{eq-identity}), the first term
vanishes because $\Tr(J)=0$. Moreover, $\Tr(b_kJ)=-2\,\delta_{k,3}$ 
so that

$$
\|T w\|^2
\;=\;
\sum_{j=1}^3\,(\Ad_T^\Bb)_{3,j}\,
(-1)\det(b_j)\,\langle w|Jb_j|w\rangle
\mbox{ . }
$$

\noindent Let us define a map $g:\CC^2\to\RR^3$ by

$$
g(w)
\;=\;
-\left(
\begin{array}{c}
\det(b_1)\,\langle w|Jb_1|w\rangle
\\
\det(b_2)\,\langle w|Jb_2|w\rangle
\\
\det(b_3)\,\langle w|Jb_3|w\rangle
\end{array}
\right)
\;=\;
\left(
\begin{array}{c}
w_1\overline{w_2}+\overline{w_1}w_2
\\
|w_2|^2-|w_1|^2
\\
|w_1|^2+|w_2|^2
\end{array}
\right)
\mbox{ , }
\qquad
w
\;=\;
\left(
\begin{array}{c}
w_1
\\
w_2
\end{array}
\right)
\mbox{ . }
$$

\noindent Then one has 

\begin{equation}
\label{eq-identity2}
\|T w\|^2
\;=\;
\langle \vec{e}_3|\,\Ad_T^\Bb\,|g(w)
\rangle
\mbox{ . }
\end{equation}

\noindent Since $g(v)=\vec{e}_3$, this implies in particular 

\begin{equation}
\label{eq-identity2b}
\|T v\|^2
\;=\;
\langle \vec{e}_3|\,\Ad_T^\Bb\,|\vec{e}_3\rangle
\mbox{ . }
\end{equation}

\section{Calculation of the averaged Landauer resistance}
\label{sec-land}

The Landauer resistance $\rho_\lambda(N)$ 
of a system of length $N$ is defined by

$$
\rho_\lambda(N) 
\;=\;
\EE\, 
\left(\Tr\left(|\TV_\lambda(N)|^2\right)\right)
\;=\;
2\,\EE\, 
\left(\|\TV_\lambda(N)v\|^2 \right)
\mbox{ , }
$$

\noindent with $v$ as in (\ref{eq-coeff}). Because $|\TV_\lambda(N)|^2$ is
positive and has unit determinant, one has $\rho_\lambda(N)\geq 2$ for all $N$. 
Using the identity (\ref{eq-identity2b}) and then the representation property
$\Ad_{ST}^\Bb=\Ad_S^\Bb\Ad_T^\Bb$ iteratively, 
the expectation value appearing in the Landauer resistance can readily
be calculated:

$$
\rho_\lambda(N)
\;=\;
2\;\langle \vec{e}_3|\,(\EE\,(\Ad_{T_\sigma}^\Bb))^N\,|\vec{e}_3
\rangle
\mbox{ , }
$$

\noindent so that $\hat{\gamma}(\lambda)=\lim_{N\to\infty}
\log(\rho_\lambda(N))/(2N)$. 
Replacing the normal form
(\ref{eq-normal}) (recall that the phases disappear right away in the
definition of the Landauer conductance):

$$
\rho_\lambda(N)
\;=\;
2\;
\langle (\Ad_{M^{-1}}^\Bb)^t\,\vec{e}_3|\,\left(\EE\,
\Ad_{R_\sigma\exp(\lambda P_\sigma+\lambda^2Q_\sigma+\Oo(\lambda^3))}^\Bb
\right)^N\,|\Ad_{M}^\Bb\vec{e}_3
\rangle
\mbox{ . }
$$

\vspace{.2cm}

Next we need to do (non-degenerate, in $\lambda$) perturbation theory of the
eigenvalues $\mu_1,\mu_2,\mu_3$ of
$\EE\,\Ad_{R_\sigma\exp(\lambda
P_\sigma+\lambda^2Q_\sigma+\Oo(\lambda^3))}^\Bb$, 
which according to
(\ref{eq-adjointperturb}) is
up to $\Oo(\lambda^3)$ given by

$$
\EE(\Ad_{R_\sigma}^\Bb)\;+\;
\lambda\,\EE(\Ad_{R_\sigma}^\Bb\ad^\Bb_{P_\sigma})
\;+\;\lambda^2\,\EE(\Ad_{R_\sigma}^\Bb\ad^\Bb_{Q_\sigma})
\;+\;\frac{\lambda^2}{2}
\EE(\Ad_{R_\sigma}^\Bb(\ad^\Bb_{P_\sigma})^2)
\mbox{ , }
$$

Let us note that the
eigenvalues of $\EE(\Ad^\Bb_{R_{\eta_\sigma}})$ are 
$\mu_1=\EE(e^{2\imath\eta_\sigma}),\,
\mu_2=\EE(e^{-2\imath\eta_\sigma}),\,\mu_3=1$ with the
eigenvectors $\vec{v}_1$, $\vec{v}_2$ and $\vec{v}_3$ respectively. 
Therefore, unless $\eta_\sigma$ is independent of
$\sigma$, the matrix $\EE(\Ad^\Bb_{R_{\eta_\sigma}})$ is not an element
of $\mbox{SO}(2,1)$ and two of its (complex conjugate)
eigenvalues are strictly within the
unit circle. 

\vspace{.2cm}

We first focus on the eigenvalue $\mu_3$. Because $\langle
\vec{e}_3|\ad^\Bb_P|\vec{e}_3 \rangle=0$, one has
$\mu_3=1+\Oo(\lambda^2)$ and the eigenvector is
$\vec{v}_3+\Oo(\lambda^2)$. Second order perturbation theory now shows

\begin{eqnarray}
\mu_3
& = &
1\;+\;\frac{\lambda^2}{2}\,
\langle\vec{v}_3|\,
\EE\,(\Ad_{R_\sigma}^\Bb
(\ad^\Bb_{P_\sigma})^2 )
|\vec{v}_3\rangle
\nonumber 
\\
& & \;+\;\lambda^2
\,
\langle \vec{v}_3|\EE(
\Ad_{R_\sigma}^\Bb\ad^\Bb_{P_\sigma})
\frac{1}{{\bf 1}-\EE(\Ad_{R_\sigma}^\Bb)}
\left({\bf 1}-|\vec{v}_3\rangle\langle \vec{v}_3|\right)
\EE(
\Ad_{R_\sigma}^\Bb\ad^\Bb_{P_\sigma})
|\vec{v}_3\rangle
\;+\;
\Oo(\lambda^3)
\mbox{ . }
\nonumber
\end{eqnarray}

\noindent Recalling $(b+c)^2+4a^2=4\,|\beta_\sigma|^2$ and
using $\langle \vec{v}_3|\ad^\Bb_{P_\sigma}|\vec{v}_1\rangle 
=\frac{1}{\sqrt{2}}(b+c+2\imath a)=\imath\sqrt{2}\,\beta_\sigma$, we
deduce

\begin{equation}
\label{eq-mu3asymp}
\mu_3
\;=\;
1\;+\;2\,\lambda^2\;
\left[
\EE(|\beta_\sigma|^2)\;+\;2\,\Re e\left(
\frac{ \EE(\beta_\sigma)\,
\EE(\overline{\beta_\sigma}\,e^{2\imath\eta_\sigma})}{
1-\EE(e^{2\imath\eta_\sigma})}
\right)\right]
\;+\;
\Oo(\lambda^3)
\mbox{ . }
\end{equation}

\noindent Now let us analyse the eigenvalues $\mu_1$ and
$\mu_2$. If $|\,\EE(e^{2\imath\eta_\sigma})|<1$ (as for the dimer
model), 
they are strictly
within the unit disc and remain there also for $\lambda$ sufficiently
small. As $\rho_\lambda(N)\geq 2$, we conclude that $\mu_3\geq 1$ 
(this also implies $D\geq 0$, like in Proposition \ref{prop-pos}, because
otherwise $\rho_\lambda(N)\to 0$ for
$N\to\infty$). It follows that the scaling exponent of the Landauer resistance
defined in (\ref{eq-land}) is given solely by $\mu_3$, namely
$\hat{\gamma}(\lambda)=\frac{1}{2}(\mu_3-1)+\Oo(\lambda^3)$ so that,
when comparing with (\ref{eq-lyapasymp}),
$\hat{\gamma}(\lambda)=2\,{\gamma}(\lambda)+\Oo(\lambda^3)$
as claimed in Theorem \ref{theo-main}.

\vspace{.2cm}

In the case where $\eta_\sigma=\eta$ independently of
$\sigma$ (as in the Anderson model), one has $D\geq 0$. 
First order perturbation theory (in $\lambda$)
shows that $\mu_{1,2}$ move along the the unit circle, while in second order
they lie inside of the unit circle. Thus the same argument as above applies
to deduce Theorem \ref{theo-main}.

\section{Higher orders for the Anderson model}
\label{sec-Anderson}

This section serves two purposes: it provides an example to which the
general theory applies and we moreover outline (algebra is left to the
reader) how it can be extended to
calculate the next higher order in perturbation theory. The
implications of the results have already been discussed in the
introduction. 

\vspace{.2cm}

The one-dimensional Anderson model is a random Jacobi matrix given by
the finite difference equation

$$
-\psi_{n+1}+\lambda\,v_n\, \psi_n -\psi_{n-1}\;=\;E\,\psi_n
\mbox{ . }
$$

\noindent Here $|E|<2$ is a fixed energy and
$v_n$ are centered real i.i.d. random variables with finite moments. 
This equation is rewritten as usual using transfer matrices:

$$
\left(\begin{array}{c} \psi_{n+1} \\ \psi_n \end{array}\right)
\;=\;
T_{\lambda,n}\;
\left(\begin{array}{c} \psi_{n} \\ \psi_{n-1} \end{array}\right)
\mbox{ , }
\qquad
T_{\lambda,n}\;=\;
\left(\begin{array}{cc} \lambda v_n-E & -1 \\ 1 & 0 \end{array}\right)
\mbox{ . }
$$

\noindent As above, we 
also write $v_\sigma$ for one of the random variables such that
$v_n=v_{\sigma_n}$ and $T_{\lambda,\sigma_n}=T_{\lambda,n}$.
For the basis change to the normal form of the transfer
matrix $T_{\lambda,\sigma}$, let us introduce

$$
E\;=\;-2\cos(k)
\mbox{ , }
\qquad
M\;=\;
\frac{1}{\sqrt{\sin(k)}}
\left(\begin{array}{cc} \sin(k) & 0
\\ -\cos(k) & 1 \end{array}\right)
\mbox{ . }
$$

\noindent It is then a matter of computation to verify

$$
MT_{\lambda,\sigma}M^{-1}
\;=\;
R_k(1+\lambda \,P_\sigma)
\mbox{ , }
\qquad
P_\sigma\;=\;
-\,\frac{v_\sigma}{\sin(k)}
\left(\begin{array}{cc} 0 & 0
\\ 1 & 0 \end{array}\right)
\mbox{ . }
$$

\noindent Comparing with (\ref{eq-normal}), we see that the rotation
$R_k$ by the angle $k$ is not random in this example, that $P_\sigma$ is
nilpotent so that $\exp(\lambda P_\sigma)=1+\lambda P_\sigma$ and that
$Q_\sigma=0$. Furthermore 

$$
\beta_\sigma
\;=\;
\frac{\imath\,v_\sigma}{2\,\sin(k)}
\mbox{ , }
\qquad
\langle v|P_\sigma|v\rangle
\;=\;-\beta_\sigma
\mbox{ . }
$$

\noindent One has $\beta_\sigma^2=-|\beta_\sigma|^2$,
$\beta_\sigma^4=|\beta_\sigma|^4$ and

\begin{equation}
\label{eq-PAnd}
\langle e_\theta|\tilde{P}_\sigma|e_\theta\rangle
\;=\;
\beta_\sigma\,(e^{2\imath\theta}-e^{-2\imath\theta})
\mbox{ , }
\qquad
\langle e_\theta|\,|{P}_\sigma|^2|e_\theta\rangle
\;=\;
|\beta_\sigma|^2\,(2+e^{2\imath\theta}+e^{-2\imath\theta})
\mbox{ . }
\end{equation}

\noindent As $\EE(\beta_\sigma)=0$, one can immediately deduce 
from (\ref{eq-lyapasymp}) the
well-known formula \cite{Tho,PF,Luc} for the Lyapunov exponent, namely
$\gamma(\lambda)=
\frac{1}{2}\lambda^2\,\EE(|\beta_\sigma|^2)+\Oo(\lambda^3)= 
\lambda^2\,\frac{\EE(|v_\sigma|^2)}{8\sin^2(k)}+\Oo(\lambda^3)$ as long
as $e^{2\imath j k}\neq 1$ for $j=1,2$ (the latter condition excludes
band edges and the Kappus-Wegner anomaly at the band center $E=0$).

\vspace{.2cm}

In order to calculate the 4th order of the Lyapunov exponent 
and the variance,
we need higher order expansions than (\ref{eq-phaseexpan}) and 
(\ref{eq-lyaptermexpan}). We will assume
$e^{2\imath j k}\neq 1$ for $j=1,\ldots, 4$.
After some algebra,

\begin{eqnarray}
e^{2\imath \Ss_{\lambda,\sigma}(\theta)}
& = &
\frac{\langle v|R_k(1+\lambda P_\sigma)|e_\theta\rangle}{
\langle \overline{v}|R_k(1+\lambda P_\sigma)|e_\theta\rangle}
\nonumber
\\
& &
\label{eq-phaseAnd}
\\
& = &
e^{2\imath (\theta+k)}
\Big[
1\,-\,\lambda\beta_\sigma(e^{2\imath \theta}+2+e^{-2\imath \theta})
+\lambda^2\beta_\sigma^2(e^{4\imath \theta}
+3e^{2\imath \theta}+3+e^{-2\imath \theta})+\Oo(\lambda^3)
\Big]
\mbox{ , }
\nonumber
\end{eqnarray}
\noindent and, using (\ref{eq-PAnd}),
\begin{eqnarray}
\gamma_{\lambda,\sigma}(\theta)
& = &
\log(\|(1+\lambda P_\sigma)e_\theta\|)
\nonumber
\\
& = &
\Re e \left[
\lambda\beta_\sigma e^{2\imath \theta}
\;-\;
\lambda^2\beta_\sigma^2 \left(\frac{1}{2}e^{4\imath \theta}
+e^{2\imath \theta}+\frac{1}{2}\right)
\;+\;
\lambda^3\beta_\sigma^3 \left(
\frac{1}{3}e^{6\imath \theta}
+e^{4\imath \theta}
+e^{2\imath \theta}\right)
\right.
\label{eq-gammaAnd}
\\
&  &
\;\;\;\;\;\;\;\;
\left.
+\;
\lambda^4\;\beta_\sigma^4 \left(
-\frac{1}{4}e^{8\imath \theta}
-e^{6\imath \theta}
-
\frac{3}{2}e^{4\imath \theta}
-e^{2\imath \theta}
-\frac{1}{4}\right)
\;+\;\Oo(\lambda^5)\;
\right]
\mbox{ . }
\nonumber
\end{eqnarray}

\noindent Using an argument similar to Lemma \ref{lem-osci}, we deduce
from (\ref{eq-phaseAnd}) and its square that 

$$
\EE_\nu (e^{2\imath \theta})
\;=\;
\frac{\lambda^2 \EE(|\beta_\sigma|^2)}{1-e^{-2\imath k}}
\;+\:\Oo(\lambda^3)
\mbox{ , }
\qquad
\EE_\nu (e^{4\imath \theta})
\;=\;
\frac{\lambda^2 \EE(|\beta_\sigma|^2)}{1-e^{-4\imath k}}
\;+\:\Oo(\lambda^3)
\mbox{ . }
$$

\noindent Moreover, $\EE_\nu (e^{6\imath \theta})=\Oo(\lambda^2)$
and $\EE_\nu (e^{8\imath \theta})=\Oo(\lambda^2)$.
Following the argument of Section \ref{sec-lyap}, we
therefore obtain from (\ref{eq-gammaAnd}) and $\Re e(1-e^{\imath
\varphi})^{-1}=\frac{1}{2}$ 

$$
\gamma(\lambda)
\;=\;
\frac{1}{2}\,\lambda^2\,\EE(|\beta_\sigma|^2)
\;+\;\lambda^4\,\left(\frac{3}{4}\,\EE(|\beta_\sigma|^2)^2
\;-\;\frac{1}{4}\,\EE(|\beta_\sigma|^4)\right)
\;+\;\Oo(\lambda^5)
\mbox{ . }
$$

\noindent This coincides with the fourth order contribution obtained in
\cite{Luc} by using complex energy Dyson-Schmidt variables. 

\vspace{.2cm}

In order to calculate the variance according to formula
(\ref{eq-variancedef}), one first 
needs to go through the arguments of Section \ref{sec-proofs}. One
finds

$$
\EE_2\;\sum_{m=1}^\infty
\big(e^{2\imath\theta_m}-\EE_\nu(e^{2\imath\theta})\big)
\;=\;
\frac{e^{2\imath \theta_0}}{e^{-2\imath k}-1}\;
\big(1-\lambda\beta_1(e^{2\imath \theta_0}+2+e^{-2\imath
\theta_0})\big)
\;+\;\Oo(\lambda^2)
\mbox{ , }
$$

\noindent a similar expression for the correlation sum of
$e^{4\imath\theta}$, and that $\EE_2\;\sum_{m=2}^\infty
(\gamma_m-\gamma)$ is up to $\Oo(\lambda^4)$ equal to

$$
\lambda^2\,\EE(|\beta_\sigma|^2)\;
\Re e\left[
\frac{e^{2\imath \theta_0}(1-\lambda\beta_1(e^{2\imath \theta_0}+2+e^{-2\imath
\theta_0}))}{e^{-2\imath k}-1}
\;+\;
\frac{1}{2}\;
\frac{e^{4\imath \theta_0}(1-2\lambda\beta_1(e^{2\imath \theta_0}+2+e^{-2\imath
\theta_0}))}{e^{-4\imath k}-1}
\right]
\mbox{ . }
$$

\noindent Finally, after having squared (\ref{eq-gammaAnd}),

$$
\EE_\nu(\gamma_1^2)-\gamma^2
\;=\;
\frac{1}{2}\,\lambda^2\,\EE(|\beta_\sigma|^2)\;
-\;
\frac{1}{8}\,\lambda^4\,\EE(|\beta_\sigma|^4)\;
-\;
\frac{1}{2}\,\lambda^4\,\EE(|\beta_\sigma|^2)^2
\;+\;\Oo(\lambda^5)
\mbox{ , }
$$

\noindent and 

$$
\EE_\nu \left(
2\,\gamma_1\;
\EE_2\;\sum_{m=2}^\infty
(\gamma_m-\gamma)
\right)
\;=\;
\frac{7}{8}\,\lambda^4\,\EE(|\beta_\sigma|^2)^2
\;+\;\Oo(\lambda^5)
\mbox{ . }
$$

\noindent Combining these results according to (\ref{eq-variancedef}),
we obtain

$$
\sigma(\lambda)
\;=\;
\frac{1}{2}\,\lambda^2\,\EE(|\beta_\sigma|^2)\;
\;+\;
\lambda^4\,
\left(
\frac{3}{8}\,\EE(|\beta_\sigma|^2)^2\;-\;
\frac{1}{8}\,\EE(|\beta_\sigma|^4)
\right)
\;+\;\Oo(\lambda^5)
\mbox{ . }
$$

\noindent Therefore we see that Lyapunov exponent and variance are
only equal to lowest order in perturbation theory. Finally let us also argue
that there cannot exist a universal analytic function $f$ such that
$\sigma=f(\gamma)$. Indeed, if $f(x)=f_1x+f_2 x^2+\Oo(x^3)$, then
$\sigma(\lambda)=f(\gamma(\lambda))$ for the Anderson model implies in
order $\lambda^2$ that $f_1=1$, but in order $\lambda^4$ there is already a
problem due to the prefactors of $\EE(|\beta_\sigma|^4)$.


\end{document}